\documentclass[aps,prb,twocolumn,showpacs,superscriptaddress,groupedaddress]{revtex4}  % for review and submission
\usepackage{graphicx}  % needed for figures
\usepackage{dcolumn}   % needed for some tables
\usepackage{amsmath}
\usepackage{bm}        % for math
\usepackage{amssymb}   % for math
\usepackage{color}
\newcommand{\tr}{{\rm Tr}}
\newcommand{\re}{{\rm Re}}

\begin{document}

% The following information is for internal review, please remove them for submission
%\widetext
%\leftline{Version xx as of \today}
%\leftline{Primary authors: Maicol A Ochoa}
%\leftline{To be submitted to (PRB)}
%\leftline{Comment to {\tt maicol@sas.upenn.edu} by xxx, yyy}
%\centerline{\em D\O\ INTERNAL DOCUMENT -- NOT FOR PUBLIC DISTRIBUTION}

% the following line is for submission, including submission to the arXiv!!
%\hspace{5.2in} \mbox{Fermilab-Pub-04/xxx-E}

\title{Energy distribution and local fluctuations in strongly coupled open quantum systems: The extended resonant level model}

\author{Maicol A. Ochoa}
\affiliation{Department of Chemistry, University of Pennsylvania, Philadelphia PA 19401, USA}
\author{Anton Bruch}
\affiliation{\mbox{Dahlem Center for Complex Quantum Systems and Fachbereich Physik, Freie Universit\"at Berlin, 14195 Berlin, Germany}}
\author{Abraham Nitzan}
\affiliation{Department of Chemistry, University of Pennsylvania, Philadelphia PA 19401, USA}
\affiliation{School of Chemistry, Tel Aviv University, Tel Aviv 69978, Israel}

%\input author_list.tex      % D0 authors (remove the first 3 lines
                             % of this file prior to submission, they
                             % contain a time stamp for the authorlist)
                             % (includes institutions and visitors)
%\date{\today}
\date{June 29, 2016}

\begin{abstract}
We study the energy distribution in the extended resonant level model at equilibrium.  Previous investigations [Phys. Rev. B {\bf 89}, 161306 (2014), Phys. Rev. B {\bf 93}, 115318 (2016)] have found, for a resonant electronic level interacting with a thermal free electron wide-band bath, that the expectation value for the energy of the interacting subsystem can be correctly calculated by considering a symmetric splitting of the interaction Hamiltonian between the subsystem and the bath. However, the general implications of this approach were questioned [Phys. Rev. B {\bf 92}, 235440 (2015)]. Here we show that already at equilibrium, such splitting fails to describe the energy fluctuations, as measured here by the second and third central moments (namely width and skewness) of the energy distribution. Furthermore, we find that when the wide-band approximation does not hold, no splitting of the system-bath interaction can describe the system thermodynamics. We conclude that in general no proper division subsystem of the Hamiltonian of the composite system can account for the energy distribution of the subsystem. This also implies that the thermodynamic effects due to local changes in the subsystem cannot in general be described by such splitting.
\end{abstract}

\maketitle

%\section{\label{sec:level1}First-level heading}
% sections are not used for PRL papers

\section{Introduction}
 The estimation of the performance and dynamical properties of nanoscale devices requires an accurate description of the quantum mechanical properties of the system, as well as of its interaction with the environment. The latter issue is particularly important when the system-bath interaction is strong, forbidding a perturbative treatment of the problem and invalidating standard arguments (e.g. surface-volume scaling) for focusing on properties that characterize the system itself. This ambiguity in the unique definition of the subsystem concerns both classical and quantum systems, but takes a special form in quantum thermodynamics\cite{Esposito2015,Gemmer2009,Kita2010, Kosloff2013, Levy2014}, where system-bath interaction is manifested not only in rates but also in broadening of energy levels. Studies of model systems such as a nonequilibrium spin-boson\cite{Boudjada2014, Nicolin2011}, two-level quantum heat machines\cite{Klimovsky2015, Aspuru-Guzik2015}, a quantum particle interacting with a harmonic oscillator bath\cite{Hanggi2008} and a resonant level embedded in a continuous band\cite{Ludovico2014} were used to discuss the issue.

 The resonant level model, a single electronic level connecting between two free-electron reservoirs,  can serve as a platform for the understanding of larger and more realistic systems involving electron transport.  An extended system that includes a fraction of the system-bath interaction Hamiltonian, has been used as a strategy to describe this model under slow driving\cite{Ludovico2014,Esposito2016}, in an attempt to identify the portion of energy transferred that corresponds to heat. An alternative approach to the quantum thermodynamics of the system has been formulated in terms of a renormalized spectral function\cite{Esposito2015}, leading to a promising description of the entropy production of the system under slow driving, but departing from the expected form for the energy and particle number at equilibrium.  In a recently proposed third approach\cite{Bruch2015} the evolution of the composite system as a whole is followed, and ``system thermodynamic quantifiers'' are identified as parts of the corresponding properties of the whole composite system that depend on local system parameters. Due to the strong hybridization of the single level to the fermionic bath(s), such quantifiers include contributions not only of exclusive systems variables but also contributions from the surrounding electronic baths. The corresponding composite was coined the extended resonant level.  This approach provides a consistent dynamic and thermodynamic description under slow driving that is reconcilable with the equilibrium limit. Interestingly, the investigations by Ludovico {\sl et.\ al.}\cite{Ludovico2014} as well as those by Bruch {\sl et.\ al.}\cite{Bruch2015} suggest that the equilibrium energy of the extended resonant level can be represented by the expectation value of the level energy plus half the energy contribution of the interaction term. This interesting observation raises the question whether this separation reflects a fundamental underlying principle, that is, is it possible to identify an effective (or extended) system Hamiltonian that governs system behavior and properties in a consistent manner. The details of this separation, if it does indeed exists, may depend on the model used: it has been already shown in Ref.\ \citenum{Esposito2016} that the symmetric splitting found in Refs.\ \citenum{Ludovico2014} and \citenum{Bruch2015} leads to a thermodynamically consistent definition of heat under slow driving only if one assumes the wide band limit and time independent coupling to the reservoirs. Still, the possible existence of such a consistent separation, not necessarily symmetric, is by itself of interest. We note that such splitting naturally arises in molecular dynamics simulations when considering the local energy of single atomic sites (see e.g. Eq.\ (11) in  Ref.\  \citenum{Lepri2003} and subsequent discussion; Eq.\ (2) in Ref.\  \citenum{Tang2014}  and Section 3 in Ref.\ \citenum{Liu2008}). In addition, considerations of the consequence of assigning part of the interaction energy in the definition of the system internal energy in the calculation of the heat capacity were also made in Ref.\ \citenum{Hanggi2008}, for the particular case of an individual particle interacting with a harmonic bath.

In this work, we study the equilibrium energy distribution of the extended resonant level, henceforth sometimes referred to as ``extended dot'', and compare this with the predictions made on the distribution by adopting a specific splitting of the interaction term.  We find that even in the simple case of a single resonant level interacting with a wide band bath, an effective system Hamiltonian based on symmetric splitting of the system bath interaction describes only average properties, that is first moment of thermodynamic observables, but fails to describe the observables that correspond to higher moments of the equilibrium distribution. In particular, we find that this effective Hamiltonian underestimates the energy fluctuations in the extended system and fails to  reproduce the energy of the extended resonant level beyond the wide band limit.

In Sec.\ref{sec:model} we introduce the model, in Sec. \ref{sec:fluct} we calculate the fluctuations as well as the skewness of the energy distribution (second and third central moments) and show that they cannot be accounted for by assigning part of the overall Hamiltonian to the extended dot. Next, in Sec.\ \ref{sec:NWBA} we discuss the equilibrium energy of the resonant level interacting with a bath with finite band width, and reach similar conclusions. We summarize and conclude in  Sec.\ \ref{sec:conclusions}.

\section{The model}\label{sec:model}
We consider a single electronic level interacting with a fermionic bath. The Hamiltonian $\hat H$ for this system is the sum of the independent Hamiltonians for the single level $\hat H_{D}$, the fermionic bath $\hat H_B$ and the interaction between them $\hat V$ :
\begin{align}
\label{eq:Hamiltonian}
\hat H =& \: \hat H_D+ \hat H_B + \hat V ,\\
\hat H_D =& \: \hat\varepsilon_d \: \hat d ^\dagger \hat d ,\\
\hat H_B =& \sum_k \varepsilon_k \hat c^\dagger_k \hat c_k ,\\
\hat V =& \sum_k V_k \hat d^\dagger \hat c_k + V_k^* \hat c_k^\dagger \hat d , \label{eq:Interaction}
\end{align}
where $\hat d^\dagger$ ($\hat d$) creates (annihilates) an electron in the level, $\hat c^\dagger_k$ ($\hat c_k$) creates (annihilates) an electron in state $k$ of the bath, while $\varepsilon_d$ and  $\varepsilon_k$ are the corresponding single electron energies. The parameter $V_k$ represents the strength of the system-bath interaction.

 For the statistical description of the equilibrium system defined by the Hamiltonian (\ref{eq:Hamiltonian}) we adopt a grand canonical ensemble and consider initially the grand canonical partition function
 \begin{align}
   \Xi =& \: \tr \{ e^{-\beta (\hat H - \mu \hat N)}\}, \label{eq:partitionGen}
\end{align}
 and the total grand canonical potential $\Omega_{tot}$.
\begin{align}
\Omega_{tot} =& -\frac{1}{\beta} \ln \Xi . \label{eq:omegaGen}
\end{align}
 Here $\hat N$ represents the number operator, $\mu$ the chemical potential, $\beta =(k_B T)^{-1}$,  $T$ the absolute temperature and $k_B$ the Boltzmann constant.   In the free electron model, $\Omega_{tot}$ can be explicitly calculated
\begin{align}
   \Omega_{tot} =& -\frac{1}{\beta} \int \frac{d \varepsilon}{2 \pi} \rho(\varepsilon) \ln (1+ e^{-\beta(\varepsilon - \mu)}),
\end{align}
 as an integral in energy involving the density of states $\rho(\varepsilon, \varepsilon_d, \Gamma)$.  It can be shown\cite{Bruch2015} that $\rho(\varepsilon)=\rho_{\varepsilon_d}(\varepsilon)+\nu(\varepsilon)$, where $\nu(\varepsilon)$ is the density of states of the free bath while $\rho_{\varepsilon_d}(\varepsilon)$ is a contribution arising from the dot and the dot-bath coupling. The latter is given by \cite{Bruch2015}
 \begin{align}\label{rhoNWB}
   \rho_{\varepsilon_d}(\varepsilon) =& \tilde A(\varepsilon)(1- \partial_\varepsilon \Lambda(\varepsilon))- \re G^r \partial_\varepsilon \Gamma(\varepsilon),
 \end{align}
where $\tilde A(\varepsilon)$ is the level spectral function given by 
 \begin{align}
    \tilde A(\varepsilon) =& \frac{\Gamma(\varepsilon)}{(\varepsilon - \varepsilon_d -\Lambda(\varepsilon))^2+ (\Gamma(\varepsilon)/2)^2}\, .
 \end{align}
and $\re G^r(\varepsilon)$ is the real part of the retarded Green function
\begin{align}
\re G^r(\varepsilon) =& \frac{\varepsilon - \varepsilon_d -\Lambda(\varepsilon)}{(\varepsilon - \varepsilon_d -\Lambda(\varepsilon))^2+ (\Gamma(\varepsilon)/2)^2}.
\end{align}
Here $\Gamma = 2 \pi \sum_k |V_k|^2 \delta(\varepsilon - \varepsilon_k)$ is the decay rate and $\Lambda (\varepsilon)$ is the Lamb shift.
 Consequently, we identify the $\varepsilon_d$-dependent part of the grand canonical potential $\Omega$ and notice that in the wide band approximation (WBA) $\rho_{\varepsilon_d}(\varepsilon)$ is the spectral function $ A(\varepsilon, \varepsilon_d, \Gamma)$ of the dot electrons given by  
\begin{equation}
  \label{eq:SpectFunct}
  A(\varepsilon, \varepsilon_d, \Gamma) = \frac{\Gamma}{(\varepsilon - \varepsilon_d)^2+(\Gamma/2)^2}.
\end{equation}
 Consequently, the $\varepsilon_d$-dependent part of the grand potential $\Omega_{tot}$ in the WBA is 
\begin{align}
\Omega_{\varepsilon_d}=& -\frac{1}{\beta} \int \frac{d \varepsilon}{2 \pi} A(\varepsilon, \varepsilon_d, \Gamma) \ln (1+ e^{-\beta(\varepsilon - \mu)}) \label{eq:omega}.
 \end{align}
Henceforth, only this part, Eq.\ (\ref{eq:omega}), of the grand potential will appear in our calculations, and we will omit the subscript $\varepsilon_d$ from it and from the thermodynamic functions and expectation values of operators calculated from it, keeping in mind that we always refer to the $\varepsilon_d$-dependent parts of these functions and expectation values.
 
 Note that in the WBA the integral that defines the grand potential in Eq.\ (\ref{eq:omega}) is divergent. This results from the slow decay at negative energy values of the spectral function $A$ and the asymptotic behavior of the logarithmic term ($\ln (1+ \exp\{-\beta(\varepsilon - \mu)\}) \to - \beta \varepsilon$ as $\varepsilon \to -\infty$). In order to avoid this divergence, we introduce a lower bound $M$ for the relevant energies, with $M \gg |\varepsilon_d|, \Gamma, T, \mu$ and define
\begin{equation}
  \label{eq:omegaM}
     \Omega_M = -\frac{1}{\beta} \int_{-M}^{\infty} \frac{d \varepsilon}{2 \pi} A(\varepsilon, \varepsilon_d, \Gamma) \ln (1+ e^{-\beta(\varepsilon - \mu)}).
\end{equation}
 The asymptotic behavior of the integrand in Eq.\ (\ref{eq:omega}) leads to the approximate relation $\Omega \sim \Omega_M + (2 \pi)^{-1} \int_{-\infty}^{-M} d\varepsilon \, \varepsilon A $, and we can interpret the second term as the energy contribution of the semi-infinite lower part of the bath spectrum, which is infinite and responsible of the divergence. It is easily seen that the internal energy and the particle number, derived from $\Omega_M$ and $\Omega$ approach each other as $M \to \infty$.
 \begin{align}
   \langle E \rangle_M =& \frac{\partial}{\partial \beta} \beta \Omega_M - \frac{\mu}{\beta} \frac{\partial}{\partial \mu} \beta \Omega_M = \int_{-M}^{\infty} \frac{d\varepsilon}{2 \pi} \varepsilon A(\varepsilon) f(\varepsilon),\label{eq:internalenergy}\\
   \langle N \rangle_M =&  \frac{\partial}{\partial \mu} \beta \Omega_M = \int_{-M}^{\infty} \frac{d\varepsilon}{2 \pi} A(\varepsilon) f(\varepsilon),
 \end{align}
where $f(\varepsilon)= (1+\exp(\beta(\varepsilon - \mu)))^{-1}$ is the Fermi function. In particular, $\langle N \rangle_M$ approaches the exact value $\langle N \rangle$ as $1/M$. The internal energy in Eq. \eqref{eq:internalenergy} thus depends on the choice of the lower bound. However, in the present analysis we are only interested in the dependence of the thermodynamic observables on local system properties, here the dot energy $\varepsilon_d$, and these do not diverge in absence of a lower bound and approach cutoff-independent values in the limit $M\rightarrow \infty$. For example the contribution from the semi-infinte lower part to the internal energy can be estimated 
\begin{align}
  \left| \frac{\partial}{\partial \varepsilon_d} \right. & \left. \int_{-\infty}^{-M} \frac{d \varepsilon}{2 \pi} A(\varepsilon) \varepsilon \right| \notag \\
\leq& \left| \int_{-\infty}^{-M}  \frac{d \varepsilon}{2 \pi} \frac{ 2 \Gamma (\varepsilon-\varepsilon_d)\varepsilon}{(\varepsilon-\varepsilon_d)^4} \right| \sim \frac{\Gamma}{\pi M}  \xrightarrow{M \to \infty} 0; 
\end{align}
which implies that the diverging term in the system energy does not depend on local dot properties.
 In the following we write integrals without explicitly expressing the limits of integration, keeping in mind that they refer to quantities derived from  $\Omega_M$ in Eq.\ (\ref{eq:omegaM}). At the same time,  whenever integration by parts is required, constants of integration associated with the artificial lower bound $M$ are disregarded because they decay as $1/M$ or do not depend on $\varepsilon_d$. We note that thermodynamic observables derived in this way satisfy standard thermodynamics laws\cite{Bruch2015}.

\section{Fluctuations and Energy distribution}\label{sec:fluct}

As discussed above, an interesting observation about this model is that when the metal is described in the wide-band approximation, the $\varepsilon_d$-dependent part of the energy can be identified as the energy of an effective subsystem characterized by the Hamiltonian\cite{Bruch2015}
\begin{equation}
 \hat H_{eff}= \hat H_D+(1/2) \hat V. \label{eq:Heff}
\end{equation}
 This observation was made on the average energies of the system and its environment. It leaves open the question whether $\hat H_{eff}$ has an intrinsic fundamental meaning as the subsystem Hamiltonian, or is it only $\langle \hat H_{eff}\rangle$ that happens to yield the $\varepsilon_d$-dependent part of the energy for this model.  It is also interesting to explore the possibility that such splitting (not necessarily symmetric) may lead to a consistent thermodynamic theory in more general situations. In this section we explore the static properties of the energy distribution for the composite system described in Sec. \ref{sec:model} and in particular consider higher moments of the system energy.

 We can compute the second moment of the energy distribution and therefore calculate the fluctuations with respect to its mean value, by introducing a rescaling parameter $\lambda$ in the Hamiltonian 
 \begin{align}
   \hat H (\lambda) =& \lambda ( \hat H_D + \hat V + \hat H_B ),\label{eq:Hlamtot}
 \end{align}
with the consequent rescaling of the partition function $\Xi(\lambda) = \tr \{ e^{-\beta(\lambda \hat H - \mu \hat N)}\}$ and grand canonical potential $\Omega(\lambda)=-\beta^{-1}\ln \Xi(\lambda)$. As illustrated in Appendix \ref{ap:Eqfluc}, rescaling in the Hamiltonian amounts to rescaling $A(\varepsilon)$.  The energy fluctuation for the extended dot is obtained by differentiation of the grand canonical potential
 \begin{align}
   \langle \hat H^2 \rangle - \langle \hat H \rangle^2 =& -\frac{1}{\beta} \left. \frac{\partial^2}{\partial \lambda^2} \Omega \right|_{\lambda=1} \label{eq:flucThermoAppendix}
\end{align}
which can be computed after one makes the observation that 
\begin{equation}
    \frac{\partial}{\partial \lambda } A = - \Gamma \frac{\partial }{\partial \varepsilon} \re G^r - \varepsilon_d \frac{\partial }{\partial \varepsilon} A \label{eq:Afinalap1}.
\end{equation}
 Taking into account only the $\varepsilon_d$-dependent part of the grand canonical potential in Eq.\ (\ref{eq:omega}) we obtain 
\begin{align}
  \langle \hat H^2 \rangle - \langle \hat H \rangle^2 =&\int \frac{d\varepsilon}{2 \pi} \varepsilon^2 A(\varepsilon) f(\varepsilon)(1-f(\varepsilon)) . \label{eq:flucThermo}
 \end{align}
 
In a similar fashion, one can determine the energy fluctuations for a subsystems associated with part of the Hamiltonian. To this end we use the rescaled Hamiltonian
\begin{equation}
  \label{eq:Hamiltonian1}
  \hat H (\lambda_D, \lambda_{B}, \lambda_{V})= \lambda_D \hat H_D + \lambda_{B} \hat H_{B}+ \lambda_{V} \hat V .
\end{equation}
Using this in Eqs.\ \eqref{eq:partitionGen} and \eqref{eq:omegaGen} readily yields
\begin{equation}
  \label{eq:Fluctuation1}
  -\left. \frac{1}{\beta} \frac{\partial^2}{\partial \lambda_i ^2} \Omega \right|_{{\bf \lambda} \to 1 }= \langle \hat H_i^2 \rangle - \langle \hat H_{i} \rangle^2 .
\end{equation}
Parameters in the spectral function $A$ defined in Eq.\ (\ref{eq:SpectFunct}) change accordingly, i.e. $A \to A(\varepsilon, \lambda_D \varepsilon_d, \lambda_B^{-1} \lambda_V^2 \Gamma)$ (see Appendix \ref{ap:Eqfluc}). Direct computation yields 
\begin{align}
  \frac{\partial}{\partial \lambda_D} A =& - \varepsilon_d \frac{\partial}{\partial \varepsilon} A,\\
  \frac{\partial}{\partial \lambda_B} A =& \lambda_B^{-2} \lambda_V^2 \Gamma \frac{\partial}{\partial \varepsilon} \re G^r,\\
 \frac{\partial}{\partial \lambda_V} A =& - 2 \lambda_B ^{-1} \lambda_V \Gamma \frac{\partial}{\partial \varepsilon} \re G^r .
\end{align}
 Equations \eqref{eq:omegaM} and \eqref{eq:Fluctuation1} then lead to 
 \begin{equation}
   \label{eq:flucHd}
   \langle H_D^2 \rangle - \langle H_D \rangle^2 = \varepsilon_d^2 \int \frac{d\varepsilon}{2 \pi} A(\varepsilon) f(\varepsilon)(1- f(\varepsilon)).
 \end{equation}
 As discussed above and in Ref.\ \citenum{Bruch2015}, the average thermodynamic properties of the extended dot subsystem can be accounted for in this model by assigning to it the effective Hamiltonian  $\hat H_{eff}$ defined in Eq.\ (\ref{eq:Heff}), corresponding to a symmetric splitting of the interaction Hamiltonian between system and environment. Such symmetric splitting was also found to lead to a consistent heat current under {\sl ac} driving of the dot level \cite{Ludovico2014}.  Next we check if fluctuations in the energy derived from $\hat H_{eff}$ are equivalent to those given by Eq.\ (\ref{eq:flucThermo}) as far as their dependence on $\varepsilon_d$ is concerned. To this end, we adopt a rescaling of the form
\begin{equation}
\label{eq:Hamiltonian2}
\hat H  (\lambda_{eff}, \lambda_{B}, \lambda_{V}') = \lambda_{eff} \hat H_{eff} + \lambda_B \hat H_B + (1/2) \lambda_V' \hat V ,
\end{equation}
and find that parameters in the spectral function change as $A = A(\varepsilon, \lambda_{eff} \varepsilon_d, \lambda_B^{-1}(\lambda_{eff}+\lambda_V')^2(1/4)\Gamma) $. In addition the identity
\begin{equation}
\frac{\partial A}{\partial \lambda_{eff}}= - \lambda_B^{-1}\Gamma\frac{(\lambda_{eff}+\lambda_V')}{2}\frac{\partial}{\partial \varepsilon} \re G^r - \varepsilon_d \frac{\partial A}{\partial \varepsilon} \label{eq:derA2} ,
\end{equation}
holds.  Implementing Eq.\ (\ref{eq:Fluctuation1}) for this choice leads to 
\begin{align}
\langle (\hat H_{eff})^2 \rangle & -\langle \hat H_{eff} \rangle^ 2 = \notag \\
\int \frac{d \varepsilon}{2 \pi} \varepsilon^2 A(\varepsilon)& f(\varepsilon)(1 - f(\varepsilon))-\frac{1}{2 \beta} \int \frac{d\varepsilon}{2 \pi} (\varepsilon - \varepsilon_d) A(\varepsilon) f(\varepsilon) \label{eq:flucHdV}.
\end{align}

 If the Hamiltonian $\hat H_D$ was a consistent choice for the extended dot Hamiltonian when coupled to its environment, the $\varepsilon_d$ dependence of Eqs.\ (\ref{eq:flucThermo}) and (\ref{eq:flucHd})  (i.e. their derivatives with respect to $\varepsilon_d$) should have been the same. Similarly, if $\hat H_{eff}$ of Eq.\ (\ref{eq:Heff}) was such a choice, the $\varepsilon_d$ dependence of Eqs.\ (\ref{eq:flucThermo}) and (\ref{eq:flucHdV}) would have been the same.  Writing the difference between the Eq.\ (\ref{eq:flucThermo}) and Eq.\ (\ref{eq:flucHdV}) as a function of the level energy
\begin{equation}
  \label{eq:diffluct}
  g_2(\varepsilon_d)=\frac{1}{2 \beta} \int \frac{d\varepsilon}{2 \pi} (\varepsilon - \varepsilon_d) A(\varepsilon) f(\varepsilon),
\end{equation}
and calculating its derivative respect to $\varepsilon_d$\footnote{The derivative is taken in order to focus on the part of this difference that is associated with the extended dot and to discard parts that are independent of $\varepsilon_d$ and thus irrelevant for the description of the extended dot. The derivative would be zero if the presence of the dot had the same effect on the fluctuations described by Eqs.\ \eqref{eq:flucThermo} and \eqref{eq:flucHdV}}
\begin{equation}
  \label{eq:epsdiffluc}
  \frac{\partial g_2(\varepsilon_d)}{\partial \varepsilon_d} = \frac{1}{2}\int \frac{d \varepsilon}{2 \pi}(\varepsilon_d -\varepsilon) A(\varepsilon) f(\varepsilon)(1-f(\varepsilon)),
\end{equation}
we find that the effective Hamiltonian $\hat H_{eff}$ predicts a different behavior in the fluctuations upon changes in local parameters of the extended dot.

The discrepancy between the thermodynamic energy distribution of the extended dot, as described by the grand potential Eq.\ \eqref{eq:omega}, and the one of the effective Hamiltonian Eq.\ \eqref{eq:Heff}, appears also in higher orders moments of the energy distribution. For example, the dependence on $\varepsilon_d$ of the third moment (skewness) for the extended dot can be calculated using the rescaling for the Hamiltonian in Eq.\ (\ref{eq:Hlamtot}) and by differentiation respect to $\lambda$ of the rescaled grand potential in Eq.\ (\ref{eq:omegaGen})
\begin{align}
\Big \langle (\hat H -\langle \hat H \rangle )^3 \Big \rangle \notag =& \:  \left. \frac{1}{\beta^2}\frac{\partial^3 \Omega}{\partial \lambda^3} \right|_{{\bf \lambda} \to 1},
\end{align}
and in terms of the $\varepsilon_d$-dependent part of the grand potential in Eq.\ (\ref{eq:omega})
\begin{equation}
\label{eq:skewness1}
  \Big \langle (\hat H -\langle \hat H \rangle )^3 \Big \rangle = \int \frac{d \varepsilon}{2\pi} \varepsilon^3 A(\varepsilon) f(\varepsilon)(1- f(\varepsilon))(1- 2f(\varepsilon)) . 
\end{equation}
This result can be compared to that obtained from the third moment of the energy distribution associated with the ``effective dot Hamiltonian''  Eq.\ (\ref{eq:Heff}). The latter is obtained  using the rescaling for the Hamiltonian in Eq.\ (\ref{eq:Hamiltonian2}) and by differentiation respect to $\lambda_{eff}$
 \begin{align}
  \Big \langle \left(\hat H_{eff} - \langle \hat H_{eff} \rangle \right)^3 \Big \rangle =&  \left. \frac{1}{\beta^2}\frac{\partial^3 \Omega}{\partial \lambda_{eff}^3} \right|_{{\bf \lambda} \to 1}  \notag \\
= \int \frac{d\varepsilon}{2 \pi} \varepsilon^3 A(\varepsilon)& f(\varepsilon)(1 - f(\varepsilon))(1 - 2 f(\varepsilon)) \notag\\
-\frac{3}{2 \beta} \int \frac{d\varepsilon}{2 \pi} &\varepsilon (\varepsilon - \varepsilon_d) A(\varepsilon) f(\varepsilon)(1 - f(\varepsilon)). \label{eq:skewness2B}
 \end{align}

Once again, direct comparison between Eqs.\ (\ref{eq:skewness1}) and (\ref{eq:skewness2B}) demonstrates that the effective Hamiltonian $\hat H_{eff}$ does not constitute a consistent choice for the extended dot Hamiltonian and, in fact, the difference $g_3(\varepsilon_d)$ of the third moment of the energy of the extended dot and the one of the effective Hamiltonian and its derivative respect to $\varepsilon_d$ 
\begin{align}
  g_3(\varepsilon)=& \frac{3}{2 \beta} \int \frac{d\varepsilon}{2 \pi} \varepsilon (\varepsilon - \varepsilon_d) A(\varepsilon) f(\varepsilon)(1 - f(\varepsilon))\\
\frac{\partial}{\partial \varepsilon_d} g_3(\varepsilon) =& \frac{3}{2 \beta} \int \frac{d\varepsilon}{2 \pi} (\varepsilon - \varepsilon_d) A(\varepsilon) f(\varepsilon)(1 - f(\varepsilon)) \notag \\
- \frac{3}{2} \int& \frac{d\varepsilon}{2 \pi} \varepsilon (\varepsilon - \varepsilon_d) A(\varepsilon) f(\varepsilon)(1 - f(\varepsilon))(1- 2 f(\varepsilon)),
\end{align}
 reveal that upon driving in the level energy, the $\varepsilon_d$-dependent part of the skewness is incorrectly predicted by $\hat H_{eff}$.

\section{Energy Splitting beyond the wide band limit}\label{sec:NWBA}
 The effective Hamiltonian in Eq.\ \eqref{eq:Heff} was found to correctly represent the dependence of the average system energy on $\varepsilon_d$ in the wide band approximation. We now consider the extended resonant level model when this approximation regarding the bath is relaxed. In this case the retarded self energy of the dot electrons becomes a complex function of the energy, with a finite real part (Lamb shift $\Lambda(\varepsilon)$) and an energy dependent imaginary part, that is the energy dependent decay rate $\Gamma$. 
The  $\varepsilon_d$-dependent part of the grand potential reads
 \begin{align}
   \tilde \Omega =&- \frac{1}{\beta} \int \frac{d\varepsilon}{2 \pi} \rho_{\varepsilon_d}(\varepsilon) \ln (1 + e^{-\beta(\varepsilon - \mu)})\,, \label{eq:omegaNWBA}
 \end{align}
with $\rho_{\varepsilon_d}(\varepsilon)$ given by Eq.\ (\ref{rhoNWB}), setting the form of the complete equilibrium thermodynamics of the extended resonant level.
 In particular,  the $\varepsilon_d$-dependent part of the internal energy $E$ can be calculated from $\tilde \Omega$ as follows
 \begin{align}\label{ENWBA}
 E =  &\left(\frac{\partial}{\partial \beta}- \frac{\mu}{\beta}\frac{\partial}{\partial \mu} \right)\beta \tilde \Omega =\int \frac{d\varepsilon}{2 \pi} \varepsilon \rho_{\varepsilon_d}(\varepsilon) f(\varepsilon)\,.
 \end{align}
The important observation (see Appendix \ref{ap:ImportantRelation}) 
  \begin{align}\label{eq:ImportantObserv}
  \frac{\partial}{\partial \varepsilon_d} \rho_{\varepsilon_d} (\varepsilon) =& -\frac{\partial}{\partial \varepsilon} \tilde A(\varepsilon),
  \end{align}
  has the consequence that the quasistatic work
  \begin{align}
  d W  =& d\varepsilon_d \frac{\partial }{\partial \varepsilon_d} \tilde \Omega =  d\varepsilon_d \int \frac{d\varepsilon}{2 \pi}\tilde A(\varepsilon) f(\varepsilon), 
  \end{align}
  connects correctly to the force experienced by external driving also beyond the WBA. That can be seen by considering that the time dependent dot level is associated with some external coordinate, in which case the quasistatic work is the work done by the external coordinate against the quasistatic part of the adiabatic reaction forces generated by the coupling to the electronic system. \cite{BodePRB,bode2012current}
\begin{center}
  \begin{figure}[t]
    \centering
    \includegraphics[scale=0.7]{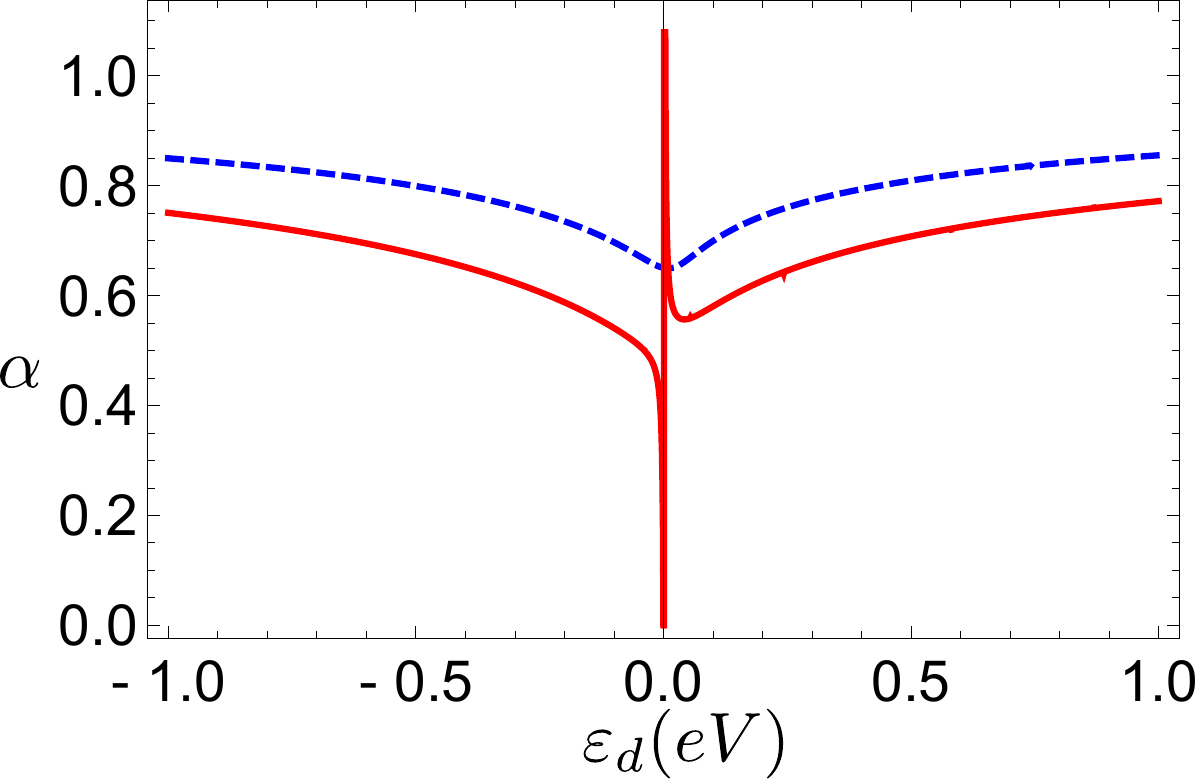}
    \caption{(Color online) Splitting factors $\alpha_1$  (blue, dashed) and $\alpha_2$  (red, solid) defined in Eqs.\ (\ref{eq:lambda}) and (\ref{eq:lambda2}) respectively, as a function of the energy level $\varepsilon_d$ for a resonant level with Lorentizian decay rate $\Gamma$ and the corresponding Lamb shift (Eqs.\ (\ref{eq:GammaL}) and (\ref{eq:LambdaL})).  Parameters for this model are: $\Gamma_o=0.1 eV$, $\mu = 0$, $W = 0.5 eV$, $E_B= 0.2 eV$, $T= 10 K$.  \label{fig:lambda}}
  \end{figure}
\end{center}
  Next we address the question whether in this situation beyond the WBA a splitting of the interaction Hamiltonian between effective bath and effective system can properly account for the internal energy of the extended resonant level.
If some consistent, not necessarily symmetric, splitting exists, then we can reproduce this energy as expectation value of the effective Hamiltonian $\langle \hat H_D + \alpha_1 \hat V \rangle = E$. Using Eq. \eqref{ENWBA} for $E$ and solving for $\alpha_1$ yields  
\begin{align}
  \alpha_1  =& \frac{\int \frac{d\varepsilon}{2 \pi} \varepsilon \rho_{\varepsilon_d}(\varepsilon) f(\varepsilon)- \langle \hat H_D \rangle}{\langle \hat V \rangle}\, , \label{eq:lambda}
\end{align}
where the resulting $\alpha_1$ should be constant ($\varepsilon_d$ independent). Alternatively, the validity of the splitting would be implied by a weaker criterion - that the dependence on $\varepsilon_d$ of the averaged effective Hamiltonian and of $E$ (Eq.\ \eqref{ENWBA}) are the same. This implies   
\begin{equation}
  \frac{\partial}{\partial \varepsilon_d}\langle \hat H_D\rangle + \alpha_2  \frac{\partial}{\partial \varepsilon_d} \langle \hat V \rangle =  \frac{\partial}{\partial \varepsilon_d} E.
\end{equation}
 This leads to \footnote{Note that the attempt to reproduce the quasistatic heat current leaving the extended dot via the energy flow into the effective bath $ \hat H_{B}+(1-\alpha_2) \hat V$ leads to the same equation for $\alpha_2$ Eq. \eqref{eq:lambda2}. \cite{Bruch2015} }
\begin{equation}
  \label{eq:lambda2}
  \alpha_2 = \frac{ \frac{\partial}{\partial \varepsilon_d}\int \frac{d\varepsilon}{2 \pi} \varepsilon \rho_{\varepsilon_d}(\varepsilon) f(\varepsilon) -  \frac{\partial}{\partial \varepsilon_d}\langle \hat H_D\rangle}{\frac{\partial}{\partial \varepsilon_d} \langle \hat V \rangle}.
\end{equation}
 Again, if splitting works, the resulting $\alpha_2$ would be a constant, independent of $\varepsilon_d$.  The expectation values of $\hat H_D$ and $\hat V$ can be either calculated from  the grand potential $\tilde \Omega$ as described in Appendix \ref{ap:ImportantRelation} and \ref{ap:energyV} respectively, or directly by computing $\langle \hat H_D \rangle$ and $\langle \hat V \rangle$ within the Green function formalism. They take the form
 \begin{align}
   \langle \hat H_D \rangle =& \varepsilon_d \int \frac{d\varepsilon}{2 \pi } \tilde A (\varepsilon) f(\varepsilon) \label{eq:HdNWBA},\\
\langle \hat V \rangle =& 2 \int \frac{d\varepsilon}{2 \pi }(\varepsilon - \varepsilon_d) \tilde A (\varepsilon) f(\varepsilon) \label{eq:VNWBA}.
 \end{align}
In the wide band limit, $\rho_{\varepsilon_d}(\varepsilon) \to A(\varepsilon)$ and $\tilde A (\varepsilon) \to A(\varepsilon)$ leads to  $\alpha_1 \to 1/2$ in Eq. \eqref{eq:lambda} (which can be used to show that also $\alpha_2 \to 1/2$), independent of local parameters.  Figure \ref{fig:lambda} shows the splitting factors $\alpha_1$ and $\alpha_2$ calculated from Eqs.\ \eqref{eq:lambda} and \eqref{eq:lambda2} plotted against $\varepsilon_d$, for a model with a Lorentzian form of the decay rate. For this model the Lamb shift is given by 
\begin{align}
  \Gamma (\varepsilon) =& \Gamma_o \frac{W^2}{W^2+(\varepsilon - E_B)^2} \label{eq:GammaL},\\
  \Lambda (\varepsilon) =& \frac{\Gamma_o}{2} \frac{W(\varepsilon - E_B)}{W^2+(\varepsilon - E_B)^2} \label{eq:LambdaL},
\end{align}
where $W$  and $E_B$  are the width and the center of the band, respectively, and $\Gamma_o$ is the decay rate at the center of the band.
Clearly, the symmetric splitting suggested in the WBA fails to predict the $\varepsilon_d$ dependence of the system energy. Moreover, the fact that the calculated splitting parameters depends on the dot level $\varepsilon_d$  implies that there does not exist a splitting factor that can be used to write an effective dot Hamiltonian in the general non-wide-band model.

\section{Conclusions}\label{sec:conclusions} 
 For the resonant level model, Eqs.\ (\ref{eq:Hamiltonian})-(\ref{eq:Interaction}), splitting the system-bath interaction symmetrically and taking Eq.\ (\ref{eq:Heff}) to represent the system Hamiltonian has been useful in analyzing the average thermal properties of this model\cite{Ludovico2014,Bruch2015} in the wide band approximation. The present analysis indicates that this symmetric splitting does not reflect any fundamental physics and fails when considering higher moments of the energy distribution even in the wide-band limit. In particular, we observe that energy fluctuations and the asymmetry of the distribution are incorrectly estimated by this choice. Therefore using the fluctuation-dissipation theorem for evaluating transport coefficients with fluctuations in properties of the effective system should be done with care.  Consistent equilibrium thermodynamics for the strongly coupled resonant level model can be extended to situations beyond the WBA. However, simple representation of the dependence of the system internal energy on local dot parameters in terms of the expectation value of an effective system Hamiltonian that splits the coupling Hamiltonian between system and bath does not generally exist. The correct energy distribution at equilibrium and its dependence on local dot properties can be obtained outside the WBA only by studying the full system.

\section{Acknowledgements}
This work was supported in part by the Israel Science Foundation and the USA-Israel binational Science Foundation (AN), the SFB 658 of the Deutsche Forschungsgemeinschaft (AB) and the University of Pennsylvania (MAO, AB and AN)

\appendix

\section{Derivation of Eqs.\ \ref{eq:flucThermoAppendix} and \ref{eq:flucThermo}}\label{ap:Eqfluc}
In this appendix we derive the expression for the local energy fluctuations of the extended dot.  We consider the rescaled Hamiltonian in Eq.\ (\ref{eq:Hlamtot}) and grand canonical potential in Eq.\ \eqref{eq:omegaGen}, with rescaling parameter $\lambda$ and evaluate
\begin{align}
-  \frac{1}{\beta} \frac{\partial^2}{\partial \lambda^2} \Omega(\lambda) &=-\frac{1}{\beta^2}\left( \frac{1}{\Xi}\frac{\partial \Xi}{\partial \lambda} \right)^2 +\frac{1}{\beta^2}\frac{1}{\Xi(\lambda)}\frac{\partial^2 \Xi(\lambda)}{\partial \lambda^2}
\end{align}
Setting $\lambda=1$ we obtain Eq.\ (\ref{eq:flucThermoAppendix}). Next, we notice that the rescaled Hamiltonian has effective level energy $\lambda \varepsilon_d$, system-bath coupling parameter $\lambda V_k$ and bath electron energies $\lambda \varepsilon_k$. Accordingly, the rescaled electron decay rate $\tilde \Gamma$ in the WBA is $\tilde \Gamma = 2 \pi \sum_k |\lambda V_k|^2 \delta(\varepsilon - \lambda \varepsilon_k)= 2 \pi \lambda^2 \times (\lambda^{-1}) \sum_k |V_k|^2 \delta(\lambda^{-1}\varepsilon - \varepsilon_k) =\lambda \Gamma $, and the level spectral function depends on $\lambda$ as follows
\begin{equation}
  A = \frac{\lambda \Gamma}{(\varepsilon-\lambda \varepsilon_d)^2+(\lambda \Gamma/2)^2}.
\end{equation}
Evaluating the derivatives of $A$ with respect to $\lambda$, as well as the derivatives of $A$ and $\re G^r$ with respect to the energy $\varepsilon$ we obtain  Eq.\ \eqref{eq:Afinalap1}.
%Denoting by $B=B(\varepsilon, \lambda)=(\varepsilon - \lambda \varepsilon_d)^2 +(\lambda \Gamma/2)^2$ we find
%\begin{align}
% \frac{\partial}{\partial \lambda } A =& \frac{\Gamma}{B^2} \left(( \varepsilon - \lambda \varepsilon_d)^2-(\lambda \Gamma/2)^2\right) +  \varepsilon_d 2 \lambda\frac{ \Gamma}{B^2} (\varepsilon -\lambda \varepsilon_d). \label{eq:Alambdaap1}
%\end{align}
%he energy derivatives of $A$ and $\re G^r$
%begin{align}
% \frac{\partial }{\partial \varepsilon } A =& -2\lambda \frac{\Gamma}{B^2}(\varepsilon - \lambda \varepsilon_d), \label{eq:Adeap1}\\
%frac{\partial }{\partial \varepsilon } \re G^r =& -\frac{1}{B^2}\left( (\varepsilon - \lambda \varepsilon_d )^2-(\lambda \Gamma/2)^2\right),\label{eq:ReGrap1}
%end{align}
%an be substituted into Eq.\ (\ref{eq:Alambdaap1}) to obtain Eq.\ (\ref{eq:Afinalap1}).
 Also, Eqs.\ (\ref{eq:omegaM}), (\ref{eq:flucThermoAppendix}) and (\ref{eq:Afinalap1}) readily yield
\begin{align}
 \left. \frac{\partial^2}{\partial \lambda^2} \Omega \right|_{\lambda=1}=& \int \frac{d\varepsilon}{2 \pi} \varepsilon^2 A \frac{\partial}{\partial \varepsilon} f(\varepsilon)\label{eq:apderfluc},
\end{align}
which is essentially Eq.\ (\ref{eq:flucThermo}). 

 More generally, the above discussion shows that by identifying how the parameters of the Hamiltonian change after rescaling, we can determine the functional form of the spectral function $A$ on the rescaling parameters $\lambda_i$, then use its derivatives with respect to these scaling parameters to find averages and higher moments of other relevant local quantities expressed in terms of energy derivatives of $A$ and $\re G^r$, and the resulting expression used in the computation of local quantities.

\section{Derivation of Eqs.\ \ref{eq:ImportantObserv}  and \ref{eq:HdNWBA} }\label{ap:ImportantRelation}
Here we derive the relation $\frac{\partial}{\partial \varepsilon_d} \rho_{\varepsilon_d} (\varepsilon) = -\frac{\partial}{\partial \varepsilon} \tilde A(\varepsilon)$ used in Sec. \ref{sec:NWBA}.
Let $B(\varepsilon) = (\varepsilon-\varepsilon_{d}-\Lambda)^{2}+(\Gamma/2)^{2}$, such that $\tilde A = \Gamma / B$ and $\re G^r = (\varepsilon-\varepsilon_{d}-\Lambda)/B $. Thus 
\begin{equation}
\frac{\partial}{\partial \varepsilon_d} \tilde A = \frac{1}{B^2} 2\left(\varepsilon-\varepsilon_{d}-\Lambda\right)\Gamma\label{dA},
\end{equation}
\begin{equation}
	\frac{\partial}{\partial \varepsilon_{d}} \re G^r  = \frac{1}{B^2}\left\{\left(\varepsilon-\varepsilon_{d}-\Lambda\right)^{2}-\left(\Gamma/2\right)^{2}\right\}\label{dReGr}.
\end{equation}
The energy derivative of the extended dot spectral function $\tilde A$ is
\begin{align}
	\frac{\partial}{\partial \varepsilon} \tilde A  = & \frac{1}{B^2}\left\{ \left(\left(\varepsilon-\varepsilon_{d}-\Lambda\right)^{2}+(\Gamma/2)^{2}\right) \partial_{\varepsilon}\Gamma \right. \notag\\
	 &\left.-\left(2\left(\varepsilon-\varepsilon_{d}-\Lambda\right)\left(1-\partial_{\varepsilon}\Lambda\right)\Gamma+2(\Gamma/2)^2 \partial_{\varepsilon}\Gamma \right)\right\}\label{eq:B3}\\
	 = & \partial_{\varepsilon}\Gamma\frac{\partial}{\partial \varepsilon_{d}} \re G^r-\left(1-\partial_{\varepsilon}\Lambda\right)\frac{\partial}{\partial\varepsilon_{d}}\tilde A\\
	 =&- \frac{\partial}{\partial \varepsilon_d} \rho_{\varepsilon_d} (\varepsilon)\,,
\end{align}
where we used Eq. \eqref{dA} and  \eqref{dReGr} in \eqref{eq:B3} in order to identify $ \rho_{\varepsilon_d} (\varepsilon)$ as given by Eq. \eqref{rhoNWB}.

 To obtain $\langle \hat H_D \rangle$ in Eq.\ (\ref{eq:HdNWBA}), just notice that $\langle \hat H_D \rangle = \varepsilon_d \langle \hat d^\dagger \hat d \rangle =\varepsilon_d \partial_{\varepsilon_d} \tilde \Omega  =\varepsilon_d \int \frac{d\varepsilon}{2 \pi} \tilde A f(\varepsilon)$ , where we have used \eqref{eq:ImportantObserv}.

\section{Derivation of Eq.\ \ref{eq:VNWBA}}\label{ap:energyV}
To obtain the expression for $\langle \hat V \rangle$ in Eq.\ (\ref{eq:VNWBA}) we defined the rescaled Hamiltonian
\begin{equation}
\hat H (\lambda) = \hat H_D + \lambda \hat V + \hat H_B,
\end{equation}
and observe that $\Gamma$  and $\Lambda$ rescale as $\Gamma = \lambda^2 \Gamma$ and $\Lambda = \lambda^2 \Lambda $, respectively. The rescaled retarded Green function and spectral density are
\begin{align}
  \label{eq:ReGr2}
 \re G^r(\varepsilon, \varepsilon_d, \Gamma, \Lambda; \lambda)=& \frac{\varepsilon- \varepsilon_d-\lambda^2 \Lambda}{(\varepsilon- \varepsilon_d-\lambda^2 \Lambda)^2 + (\lambda^2  \Gamma/2)^2}\\
  \label{eq:A2}
  \tilde A(\varepsilon; \varepsilon_d, \Gamma, \Lambda; \lambda)=& \frac{\lambda^2 \Gamma}{(\varepsilon- \varepsilon_d-\lambda^2 \Lambda)^2 + (\lambda^2  \Gamma/2)^2} 
\end{align}
and the derivative of $\rho_{\varepsilon_d}$ with respect to the rescaling parameter $\lambda$ is
\begin{align}
  \frac{\partial}{\partial \lambda}\rho_{\varepsilon_d} (\varepsilon) =&  \left( \frac{\partial }{\partial \lambda}  \tilde A \right) (1-\partial_\varepsilon \Lambda) -  \tilde A(2/\lambda) \partial_\varepsilon \Lambda \notag \\
& -  \left( \frac{\partial }{\partial \lambda} \re G^r \right) \partial_\varepsilon \Gamma  -  \re G^r (2/\lambda) \partial_\varepsilon \Gamma. \label{eq:rholong}
\end{align}
 Equations \eqref{eq:ReGr2} - \eqref{eq:rholong} lead to 
\begin{align}
 \frac{\partial}{\partial \lambda}\rho_{\varepsilon_d} (\varepsilon) =&  \frac{2}{B^2} \Gamma (\varepsilon -\varepsilon_d -\Lambda)(2/\lambda) \Lambda  (1 - \partial_\varepsilon \Lambda) \notag\\
&-  (2/\lambda) \frac{\partial}{\partial \varepsilon}\left(\tilde A \Lambda \right) \notag\\ 
&+(2/\lambda) \Lambda \left(\frac{\partial}{\partial \varepsilon} \tilde A \right) \notag\\
& - (2/\lambda)\frac{\partial}{\partial \varepsilon}\left(  \Gamma\, \re G^r \right) \notag\\
&- \frac{1}{B^2}(2/\lambda) \left[ (\varepsilon -\varepsilon_d-\Lambda)^2-(\Gamma/2)^2\right]\Lambda \partial_\varepsilon \Gamma \label{eq:rhostep2}.
\end{align}
The first, third and fifth terms in the r.h.s of Eq.\ (\ref{eq:rhostep2}) mutually cancel. Therefore
\begin{equation}
  \frac{\partial}{\partial \lambda}\rho_{\varepsilon_d} (\varepsilon) =  - \frac{2}{\lambda}\left\{ \frac{\partial}{\partial \varepsilon}\left(  \Gamma\, \re G^r \right)  + \frac{\partial}{\partial \varepsilon}\left(\tilde A \Lambda \right)\right\} .\label{eq:rholong2}
\end{equation}
Finally, the expression in Eq.\ (\ref{eq:rholong2}) can be used to calculate $\langle \hat V \rangle$: 
\begin{align}
  \langle \hat V \rangle =& \left. \frac{\partial}{\partial \lambda} \tilde \Omega \right|_{\lambda=1}\\
=&-\frac{1}{\beta} \int \frac{d\varepsilon}{2 \pi} \left(  \frac{\partial}{\partial \lambda}\rho_{\varepsilon_d} (\varepsilon) \right)_{\lambda =1 } \ln \left(1+ e^{-\beta(\varepsilon-\mu)} \right)\\
=&2 \int \frac{d\varepsilon}{2 \pi} (\varepsilon -\varepsilon_d) \tilde A f(\varepsilon),
\end{align}
which is the result in Eq.\ (\ref{eq:VNWBA}). This result can also be derived from the Nonequilibrium Green Function formalism (See Ref.\ \citenum{Bruch2015}).
\bibliography{BibFile}
\end{document}